\documentclass[preprint,nofootinbib]{revtex4}

 \usepackage{amsmath} 
\usepackage{amsfonts} 
\usepackage{amssymb}
\usepackage{graphicx} 
\usepackage{float}
\makeatletter
\DeclareRobustCommand\MakeTextUppercase{%
  \@uclcnotmath{\def\i{I}\def\j{J}}{##1##2}\large}
\DeclareRobustCommand\MakeTextLowercase{%
  \@uclcnotmath{}{##2##1}\lowercase}
\protected@edef\MakeTextLowercase#1{\MakeTextLowercase{#1}}
\let\ProvidesPackage\ProvidesPackage@latex
\let\ProcessOptions\ProcessOptions@latex
\let\DeclareOption\DeclareOption@latex
\expandafter
\let\csname MakeUppercase \expandafter\endcsname
    \csname MakeTextUppercase \endcsname
\expandafter
\let\csname MakeLowercase \expandafter\endcsname
    \csname MakeTextLowercase \endcsname
\appdef\class@documenthook{%
 \switch@longtable
}%
\makeatother

\makeatletter
\relax
  \def\ps@headings{%
      \let\@oddfoot\@empty\let\@evenfoot\@empty
      \def\@evenhead{\thepage\hfil\slshape\leftmark}%
      \def\@oddhead{{\slshape\rightmark}\hfil\thepage}%
      \let\@mkboth\markboth
    \def\sectionmark##1{%
      \markboth {%
        \ifnum \c@secnumdepth >\z@
          \thesection\quad
        \fi
        ##1}{}}%
    \def\subsectionmark##1{%
      \markright {%
        \ifnum \c@secnumdepth >\@ne
          \thesubsection\quad
        \fi
        ##1}}}%
\def\ps@myheadings{%
    \let\@oddfoot\@empty\let\@evenfoot\@empty
    \def\@evenhead{\thepage\hfil\slshape\leftmark}%
    \def\@oddhead{{\slshape\rightmark}\hfil\thepage}%
    \let\@mkboth\@gobbletwo
    \let\sectionmark\@gobble
    \let\subsectionmark\@gobble
    }%
\def\ps@article{%
    \@provide\@evenhead{\let\\\heading@cr\thepage\quad\checkindate\hfil{\leftmark}}%
    \@provide\@oddhead{\let\\\heading@cr{\rightmark}\hfil\checkindate\quad\thepage}%
    \@provide\@oddfoot{}%
    \@provide\@evenfoot{}%
    \let\@mkboth\markboth
  \let\sectionmark\@gobble
  \let\subsectionmark\@gobble
}%
\def\ps@article@final{%
    \@provide\@evenhead{\let\\\heading@cr\thepage\quad\checkindate\hfil{\leftmark}}%
    \@provide\@oddhead{\let\\\heading@cr{\rightmark}\hfil\checkindate\quad\thepage}%
    \@provide\@oddfoot{}%
    \@provide\@evenfoot{}%
    \let\@mkboth\markboth
    \def\sectionmark##1{%
      \markboth{%
        \@ifnum{\c@secnumdepth >\z@}{\thesection\hskip 1em\relax}{}%
         ##1%
       }{}%
    }%
    \def\subsectionmark##1{%
      \markright {%
        \@ifnum{\c@secnumdepth >\@ne}{\thesubsection\hskip 1em\relax}{}%
         ##1%
      }%
    }%
}%
\makeatother

\makeatletter
\def\bibliographystyle{\def\@bibstyle}%
\def\bibsection{%
  \let\@hangfroms@section\@hang@froms
  \section*{\large\refname}%
  \@nobreaktrue
}%
\makeatother

\renewcommand\refname{References}

\begin{document}

\title{Origins of Mass\footnote{Invited review for the Central European Journal of Physics.   This is the supplement to my 2011 Solvay Conference talk promised there \cite{fwSolvay}.  It is adapted from an invited talk given at the Atlanta APS meeting, April 2012.}}

\author{Frank Wilczek}
\vspace*{.2in}
\affiliation{Center for Theoretical Physics \\
Department of Physics, Massachusetts Institute of Technology\\
Cambridge Massachusetts 02139 USA}
\vspace*{.3in}

\begin{abstract}
Newtonian mechanics posited mass as a primary quality of matter, incapable of further elucidation.   We now see Newtonian mass as an emergent property.  That mass-concept is tremendously useful in the approximate description of baryon-dominated matter at low energy -- that is, the standard ``matter'' of everyday life, and of most of science and engineering -- but it originates in a highly contingent and non-trivial way from more basic concepts.   Most of the mass of standard matter, by far, arises dynamically, from back-reaction of the color gluon fields of quantum chromodynamics (QCD).  Additional quantitatively small, though physically crucial, contributions come from the intrinsic masses of elementary quanta (electrons and quarks).   The equations for massless particles support extra symmetries -- specifically scale, chiral, and gauge symmetries.  The consistency of the standard model relies on a high degree of underlying gauge and chiral symmetry, so the observed non-zero masses of many elementary particles ($W$ and $Z$ bosons, quarks, and leptons) requires spontaneous symmetry breaking.   Superconductivity is a prototype for spontaneous symmetry breaking and for mass-generation, since photons acquire mass inside  superconductors.  A conceptually similar but more intricate form of all-pervasive (i.e. cosmic) superconductivity, in the context of the electroweak standard model, gives us a successful, economical account of $W$ and $Z$ boson masses.  It also allows a phenomenologically successful, though profligate, accommodation of quark and lepton masses.   The new cosmic superconductivity, when implemented in a straightforward, minimal way, suggests the existence of a remarkable new particle, the so-called Higgs particle.   The mass of the Higgs particle itself is not explained in the theory, but appears as a free parameter.   Earlier results suggested, and recent observations at the Large Hadron Collider (LHC) may indicate, the actual existence of the Higgs particle, with mass $m_H \approx 125$ GeV.   In addition to consolidating our understanding of the origin of mass, a Higgs particle with $m_H \approx 125$ GeV could provide an important clue to the future, as it is consistent with expectations from supersymmetry.  
\end{abstract}

\preprint{MIT-CTP/4379}
\maketitle

\bigskip

\section{What is Mass?}

\subsection{Newtonian Mass}

In classical physics, as epitomized in Newtonian mechanics, mass is a primary, conserved and irreducible property of matter.  Reflecting that significance, Newton spoke of mass as {\it quantity of matter} \cite{newton}.   Mass was so foundational within Newton's view of the world that he took its central feature -- what I call Newton's zeroth law of motion -- for granted, without stating it explicitly.  Newton's zeroth law of motion, which underpins use of the others, is the conservation of mass.   The Newtonian mass of a body is assumed to be a stable property of that body, unaffected by its motion.  In any collision or reaction, the sum of the masses of the incoming bodies is equal to the sum of the masses of the products; mass can be redistributed, but neither created nor destroyed.   Mass also occurs, of course, in Newton's gravitational force law.   

Newtonian mass is, in fact, {\it the\/} new primary quality of matter introduced into the foundation of classical physics.   It supplements size and shape, which the strict ``mechanical philosophy'' of Descartes and of many of Newton's scientific contemporaries had claimed should be sufficient.   

Relativity, and then quantum field theory, profoundly changed the status of mass within physics.   Both main properties of Newton's mass-concept got undermined.  In special relativity we learn that energy is conserved, but mass is not.   In general relativity we learn that gravity, in the form of space-time curvature, responds to energy, not to mass.   The word mass still appears in modern physics, and the modern usage evolved from the earlier one, but it denotes a radically different, more fluid concept.   

Though these profound changes began in earnest more than a hundred years ago, the old concept of mass remains deeply embedded in common language and in the folk physics of everyday life, not to mention in successful engineering practice.  

The demotion of mass from its position as a logical primitive in the foundation of physics challenges us to rebuild it on deeper foundations, and opens up the central question of this paper: What is the Origin of Mass?

\subsection{Relativistic Mass}

In modern physics energy and momentum are the primary dynamical concepts, while mass is a parameter that appears in the description of energy and momentum of isolated bodies \cite{wigner}.   The early literature of relativity employed some compromise definitions of mass -- specifically, velocity-dependent mass, in both longitudinal and transverse varieties.  Those notions have proved to be more confusing than useful.    They do not appear in modern texts or research work, but they persist in some popularizations, and of course in old books.   To forestall confusion and set the stage, I'll now briefly recall the pertinent facts and definitions, using exclusively the now-standard relativistic mass concept.   This will also be an opportunity to highlight an elementary but profound point that is widely overlooked.

Let us contrast the Newtonian and relativistic equations for momentum, in terms of mass $m$ and velocity $v$
\begin{eqnarray}
p_N ~&=&~ m v \ \ \ \ \ \ \ \ \ \ \ \ \ \ \ \ \  {\rm Newtonian~ inertia} \label{NewtonP} \\ 
p_R ~&=&~ m v \frac{1}{\sqrt{1 - \frac{v^2}{c^2}}} \ \ \ \ \ {\rm relativistic ~inertia} \label{EinsteinP}
\end{eqnarray} 
where $c$ is the speed of light.    $p$ is a measure of the body's resistance to acceleration, or inertia.  Both Newtonian and Einsteinian mechanics posit that different observers, who move at constant velocity relative to one another, construct equally valid descriptions of physics, using the same laws.  To two such observers the body appears to move with different velocities, and also to have different momenta, but both observers will infer the same $m$.   Similarly, for the energy we have 
\begin{eqnarray}
E_N ~&=&~ \frac{m}{2} v^2 \ \ \ \ \ \ \ \ \ \ \ \ \ \ \ \ \  {\rm Newtonian~ kinetic~ energy} \label{NewtonE} \\ 
E_R ~&=&~ m c^2 \frac{1}{\sqrt{1 - \frac{v^2}{c^2}}} \ \ \ \ \ \ {\rm relativistic ~energy} \label{EinsteinE}
\end{eqnarray} 

The relationship between Eqns.\,(\ref{NewtonP}, \ref{EinsteinP}) is straightforward: For $v \ll c$, Eqn.\,(\ref{EinsteinP}) goes over into Eqn.\,(\ref{NewtonP}).    
Not so the relationship between Eqns.\,(\ref{NewtonE}, \ref{EinsteinE}).    Expanding $E_R$ for $v \ll c$, we have approximately
\begin{equation}
E_R ~\approx~ mc^2 + \frac{m}{2} v^2 ~=~ mc^2 + E_N \label{EversusE}
\end{equation}
The first term on the right-hand side of Eqn.\,(\ref{EversusE}) is of course the famous mass-energy $E = mc^2$ associated with bodies at rest.   

Now consider two slowly-moving bodies that interact with each other weakly, so that we can neglect potential energy.   If only conservation (i.e., constancy) of the total energy is assumed,
\begin{equation} \label{2BodyEnergy}
E_{R, {\rm total}} ~\approx~ m_1c^2 + \frac{m_1}{2} v_1^2 + m_2 c^2 + \frac{m_2}{2} v_2^2
\end{equation}
then we cannot deduce that $m_1, m_2$, or even $m_1 + m_2$ is rigorously constant, nor that the Newtonian kinetic energy $\frac{m_1}{2} v^2 + \frac{m_2}{2} v^2$ is constant.  Formal hocus-pocus can't conjure up three independent conservation laws from just one!  Rather, it would be natural to expect, from Eqn.\,(\ref{2BodyEnergy}),  that small changes in $m_1$ and $m_2$ could accompany the dynamical evolution.  Alternatively: Eqn.\,(\ref{2BodyEnergy}) in itself does not explain why the Newtonian kinetic energy, which after all is the second, subdominant term in the expansion of $E_R$, should be separately conserved, even approximately.   All that we can legitimately infer is that if the bodies move slowly, with $v \ll c$ for both $v = v_1, v_2$, then $m_1 + m_2$ is approximately constant, to order $v^2/c^2$.  Indeed, if we divide Eqn.\,(\ref{2BodyEnergy}) through by $c^2$, we arrive at
\begin{equation}
E_{R, {\rm total}}/c^2 ~\approx~ (m_1 + m_2 ) (1 + {\rm order} \ \frac{v^2}{c^2}) 
\end{equation}
These difficulties of principle actually come into play in describing nuclear reactions, where neither mass nor Newtonian kinetic energy is separately conserved, even when all the bodies involved move slowly.

In the most radical departure from the Newtonian framework, we are allowed to consider bodies with zero mass.  (And, as we'll see, that possibility proves to be very fruitful.)  The momentum and energy $p_R, E_R$ can have sensible, finite limits as $m \rightarrow 0$, with $v \rightarrow c$ appropriately.   Thus isolated bodies with $m=0$ move at the speed of light, and for such bodies we have
\begin{equation}
p_R ~=~ E_R/c
\end{equation}
but no other restriction on the values of $p_R$ or $E_R$.   

These considerations sharpen the challenge of understanding the emergence of Newtonian mass as a valid approximation in the physical world.  

\subsection{Masses of Quanta}

Relativistic quantum field theory introduces a powerful constructive principle into the reductionist program \cite{qftReview}.  Since quantum fields create (and destroy) particles, space-time uniformity of the fields -- {\it i. e}. their invariance under space and time translation -- implies that their associated particles will have the same properties, independent of where and when they are observed.    Thus all electrons, for example, have the same properties, because they are all excitations of a single universal quantum field, the electron field.   

Thoughtful atomists, notably including Newton and Maxwell, were highly aware that the most elementary facts of chemistry -- that is, the exact reproducibility of chemical reactions, including their intricate specific rules of combination -- called for the building blocks of matter to have this feature of accurate sameness, or universality, across space and time.  The macroscopic bodies of everyday experience, of course, do not -- they come in different sizes, shapes, and composition, and can accrete or erode over time.   In our ordinary experience, only artfully manufactured products can approximate to uniformity.    Both Newton \cite{newtonGod} and Maxwell \cite{maxwellGod} inferred that the basic building blocks of matter were manufactured by God at the time of creation.   Modern quantum field theory opens the possibility of an alternative explanation.  

For our problem of the origin of mass, this general principle gives most welcome simplification and guidance \cite{qftReview}.   Rather than having to address the mass of each object in the universe separately, we can focus on the properties of a few quantum fields, whose excitations (quanta) are the building blocks of matter.   Thus for instance if we understand the properties (including mass) of one electron we understand the properties of all electrons.  More generally: If we understand the properties of the fields associated with the building blocks of matter, we should be able to deduce the properties -- including mass! -- of matter itself, and those deductions will be valid universally.

\section{Emergent Mass}

\subsection{Mass for Standard Matter}

We can reach a deeply penetrating -- though, as we shall see, still incomplete and in part tentative -- understanding of the origin of the (Newtonian) mass of ordinary matter following the strategy Newton called Analysis and Synthesis, or essentially, in modern language, Reductionism.  More specifically, we can build up our description of matter mathematically starting from rigorously tested properties of a few reproducible building blocks that obey ideally simple laws.   I will return to further discuss {\it why\/} this strategy works, but first let us consider, in meaningful detail, {\it how\/} it works.   

(By ordinary or standard matter I understand the matter encountered in everyday life, and studied in materials science, chemistry, biology, geology, and stellar astrophysics.   Excluded for now are the highly unstable species produced at high-energy accelerators and the astronomical dark matter and dark energy.)

I will carry out the analysis in two stages: a semi-phenomenological stage, to identify the basic issues, and then an exploration of those issues.  The first stage occupies this section of the paper; the second stage most of its remainder.   

Atoms, in the sense of modern physics and chemistry, are a convenient point of departure.   An atom consists of its nucleus, wherein is concentrated all the positive electric charge and overwhelmingly most of the mass ($> 99.9 \%$, in all cases), surrounded by a cloud of much lighter electrons.   So the first order of business is to understand the origin of the mass of atomic nuclei.   

The conventional model of nuclei is built on the idea that they are made up from protons and neutrons (i.e., nucleons).   This picture has both rigorous and approximate aspects.   The rigorous aspect is based on discrete, additive quantum numbers and very accurate conservation laws.   Thus a nucleus with electric charge $Ze$ (or alternatively, atomic number $Z$) and baryon number $A$ uniquely corresponds to $P = Z$ protons and $N = (A-Z)$ neutrons.   The approximate aspect is the statement that in nuclei the protons and neutrons retain their identity, so that it is appropriate to model the nucleus as a collection of interacting protons and neutrons.   In the context of quantum chromodynamics -- QCD -- this is a highly non-trivial issue!    For in QCD the protons and neutrons themselves are complex objects, built up from the more basic quarks and gluons.   One might have anticipated that when nucleons are brought together their quarks and gluons would undergo complete reorganization, within which the original nucleon structure would become unrecognizable.   (In the language of the bag model, the issue is why the separate nucleon membranes do not fuse, to produce one big bag \cite{bagModel}.)    Why such reorganization does not occur, is the foremost issue regarding the relationship between fundamental QCD and traditional nuclear physics.  Before addressing that issue, however, it is appropriate to recall its empirical grounding.   The simplest and most important success of the (nearly) independent nucleon model of nuclei is directly related to our central concern, namely its account of nuclear masses.   

For the ground state of nuclei with quantum numbers $(Z, A)$ we find 
\begin{eqnarray}\label{nuclearMasses}
M(Z, A) ~&=&~ Z M_p + (A-Z) M_n + \Delta(Z,A) \nonumber \\
\Delta(Z,A) ~&\ll&~ M(Z, A)
\end{eqnarray}
wherein of course $M(Z, A), M_p, M_n$ denote the nuclear, proton, and neutron masses respectively and, importantly, the {\it mass defect\/} $\Delta(Z,A)$ is much smaller than $M(Z, A)$.     Thus, roughly speaking, the mass of a nucleus is simply the sum of the masses of the protons and neutrons indicated by its charge and baryon number.   We can easily make this quantitative.  $\Delta(Z, A)$ is negative, corresponding to binding energy $E_B (Z, A) = - \Delta(Z, A) / c^2$.   Nuclear binding energies per nucleon peak at 
\begin{equation}
E_B (Z, A) / A \lesssim 10 \ {\rm MeV} 
\end{equation}
so that 
\begin{equation}\label{smallDelta}
| \Delta(Z, A) |  / M(Z, A) ~\lesssim~ 10^{-2}
\end{equation}
Thus the change in mass due to interaction, while significant and of course eminently detectable, is relatively small.   In this quantitative sense intra-nucleon forces are far more important than inter-nucleon forces for determining the mass of standard matter.   

Before leaving the subject of inter-nucleon forces, however, a few brief comments are in order.   Within the context of nucleon-based models of nuclei, the smallness of $\Delta$ can be interpreted to imply the smallness of kinetic as opposed to rest energies, and specifically $v/c \lesssim 10^{-1}$.   In this non-relativistic framework, it is natural to use ideas like effective instantaneous interactions through potentials and many-body position space wave functions.   

Within that circle of ideas, we can address the above-mentioned ``foremost issue'' from the nuclear side.   Why do nucleons, within a nucleus, stay separate?   At long range (that is, $r \lesssim 2\times 10^{-13}$ cm.) the dominant inter-nucleon force can be ascribed to pion exchange, and is attractive.   At nearer distances repulsive vector meson exchanges kick in, but once many channels contribute, this exchange picture becomes complicated and then breaks down altogether.   Phenomenologically, the inter-nucleon force features a so-called hard core repulsion at $r \lesssim .5 \times 10^{-14}$ cm. .    That repulsion, together with the effective repulsion associated with Fermi statistics, underlies the self-consistency of the nucleon model of nuclei.    Though this explanation is convincing, it is both semi-phenomenological and only semi-quantitative; a fuller, more microscopically based derivation of the foundations of nuclear physics would be important progress.   

$\Delta(Z, A)$ is well approximated using the famous ``semi-empirical mass formula'' of Bethe and von Weisz\"acker, which can be further refined to incorporate shell effects.   Altogether, then, we have a richly detailed, empirically successful framework in which nuclear masses can be understood, given the mass of protons and neutrons.  

The mass of atoms is, as mentioned previously, overwhelmingly dominated by the mass of nuclei.   The approximate framework that is broadly successful for nuclei -- that is, slowly moving independent particles (here electrons, together with a central nucleus), with interactions described by potentials -- becomes extremely precise and accurate for atoms.  (The measure of the electron's typical velocity is the fine structure constant $\frac{e^2}{4\pi} \equiv \alpha \approx \frac{1}{137}$.  For inner electrons in atoms with high $Z$ the measure is $Z\alpha$, which can approach unity.   In that case the binding energy can be a significant fraction of the {\it electron's\/} rest energy.   A relativistic treatment is then required, and has been carried through.)  The intrinsic mass of electrons is more than a thousand times smaller than that of nucleons, and the atomic analogue of the nuclear mass defect $\Delta$ is smaller still.    

Thus the problem of understanding origin of atomic masses reduces, to a good approximation, to the problem of understanding the origin of proton and neutron masses.  To achieve a superb approximation, we must also understand the origin of electron mass.  In the course of this reduction we have appealed to the smallness of $\alpha$ , and to some qualitative facts about the inter-nucleon forces;  a full understanding of the origin of the mass of atoms should encompass those facts as well.

\subsection{The Quantum Censor, Gaps, and Sequestration}

The arguments of the preceding subsection concerned the ground states of isolated nuclei and individual atoms.   Let us now consider how one builds from understanding of the origin of mass in those cases to understanding the origin of Newtonian mass for macroscopic bodies.   

Three closely related ideas, deeply rooted in quantum theory, make this transition possible.  The {\it quantum censor\/} can effectively suppress the complexity of composite systems.   When such systems are {\it sequestered}, so that they are subject only to perturbations below their {\it energy gap}, they appear as ideally simple particles, in a unique state (their ground state).    

Let me illustrate these general ideas in the case that is most central to the mass problem for standard matter, that is nuclear masses.   For simplicity of exposition, I will temporarily neglect the nuclear spin degree of freedom, and then return to discuss its inclusion.  Nuclei generally support, in addition to their ground state, many excited states.   The excited states, however, are separated by gaps, typically of order a few MeV; that is, their energy exceeds that of the ground state by at least a few MeV.    Now an MeV  is, by everyday standards, an enormous energy: it corresponds, for example, to temperature $T \sim 10^{10}$ K.   Rearrangements of the electrons surrounding a nucleus, such as occur in chemical reactions or mechanical responses, do not bring anything approaching that energy into play.   Thus the nuclei remain in their ground states, and effectively behave as ideal, structureless particles.   In particular: Although they are not isolated particles in the mathematical sense, within standard matter nuclei behave essentially as if they were, and the potential complications we mentioned earlier in connection with mass (non-) conservation do not arise.   

Nuclear excited states and internal degrees of freedom would come into play if nuclei could approach each other within the range of the strong force, i.e. $~2 \times 10^{-13}$ cm.   However the mutual repulsion of inner electron clouds, due to repulsive Coulomb forces and fermi statistics, effectively sequesters the nuclei.  If that mechanism of sequestration lapses (as it does for itinerant protons) then Coulomb repulsion between the nuclei themselves provides a further barrier.    

Isolated nuclei typically have a degenerate ground state, due to their spin.   Inside atoms, or inside matter generally, this degeneracy is broken, but gives rise to some extremely low-energy degrees of freedom and excitations.   Indeed, magnetic resonance imaging exploits the sensitivity of nuclear spin excitations to their chemical environment.   But precisely because so little energy is involved in nuclear spin interactions -- in other words, because the spins interact so feebly -- the associated transitions have only a minuscule effect on the total mass.    

Photons are another potential source of very low-energy excitations, immune to quantum censorship.   The radiation fields in standard matter, however, must arise from transitions in the nuclear or atomic (electronic) degrees of freedom.   However nuclei are inert for the reasons just discussed, while electronic process put only tiny amounts of energy into play, compared to the rest-energy of the nuclei, and so the impact of electromagnetic radiation on the mass of matter is negligible.   Weak decays and other forms of radioactivity {\it do\/} put nuclear energies of order $\Delta$ into play.   Were these effects faster and more common, they would vitiate the Newtonian mass-concept for accurate work.   

The same logic of quantum censorship, gaps, and sequestration that applies to nuclei in standard matter also applies another level down, to nucleons within nuclei.   Protons and neutrons, in our modern understanding, are complex composite objects, made from essentially massless quarks and gluons.   Their excitations -- baryonic resonances -- are many tens of MeV higher in energy, however.   Within nuclei, the hard core and fermi statistics sequester the protons and neutrons, as we discussed earlier, and so nucleons act as structureless particles, with the same (conserved) mass as they have in isolation.     Should quarks or gluons have internal structure, {\it e. g}. if they are strings, then that structure can be, and empirically it must be, rigorously censored and sequestered through similar mechanisms.

\section{``Mass Without Mass'' for Nucleons}

120 years ago Lorentz proposed his electromagnetic theory of the origin of the (then undiscovered!) electron's mass \cite{lorentz}.
According to Lorentz's idea, back-reaction of electric and magnetic fields resists accelerated motion of charges, thus giving rise dynamically to effects we call ``mass'', or inertia.  

Lorentz's idea, in its original form, no longer appears viable, for several reasons.  In modern quantum electrodynamics (QED) the electromagnetic correction to the electron's mass appears as a multiplicative factor.  It cannot, therefore, bootstrap a vanishing intrinsic mass to a nonzero value.   This no-go result arises because in the limit of zero electron mass the equations for electrons interacting with electromagnetic fields have enhanced symmetry (chiral symmetry), which forbids mass non-zero mass, even when back reaction is taken into account.   Furthermore, notoriously, the multiplicative correction factor is infinite, or rather ill-defined, and requires renormalization.  Aside from these very serious technical problems, there is the simple, overarching fact that the electron is no longer {\it sui generis}.   If we somehow succeeded in explaining the electron's mass as an electromagnetic effect, we'd immediately be faced with the challenge of accounting for the very different values of the muon's mass $m_\mu$, and the tau lepton's mass $m_\tau$\footnote{These comments are rooted in the perturbative understanding of QED, which seems appropriate to its observed weak coupling.   Nonperturbatively, both chiral symmetry breaking and nonlinear generation of a complex spectrum of masses can occur -- and in QCD, as we'll presently discuss, they do!}.  

But something very close to Lorentz's beautiful idea is realized in quantum chromodynamics (QCD), the modern theory of the strong nuclear force.   Indeed most ($\sim 95 \%$) of the mass of protons and neutrons, and thus of ordinary matter, is due to the back-reaction of color gluon fields resisting accelerated motion of the quarks and gluons inside.  Very detailed and impressive calculations, based on direct numerical solution of the equations and exploiting the full power of modern computers, stand behind those words.  
\begin{figure}[htbp] 
   \centering
   \includegraphics[scale=.5]{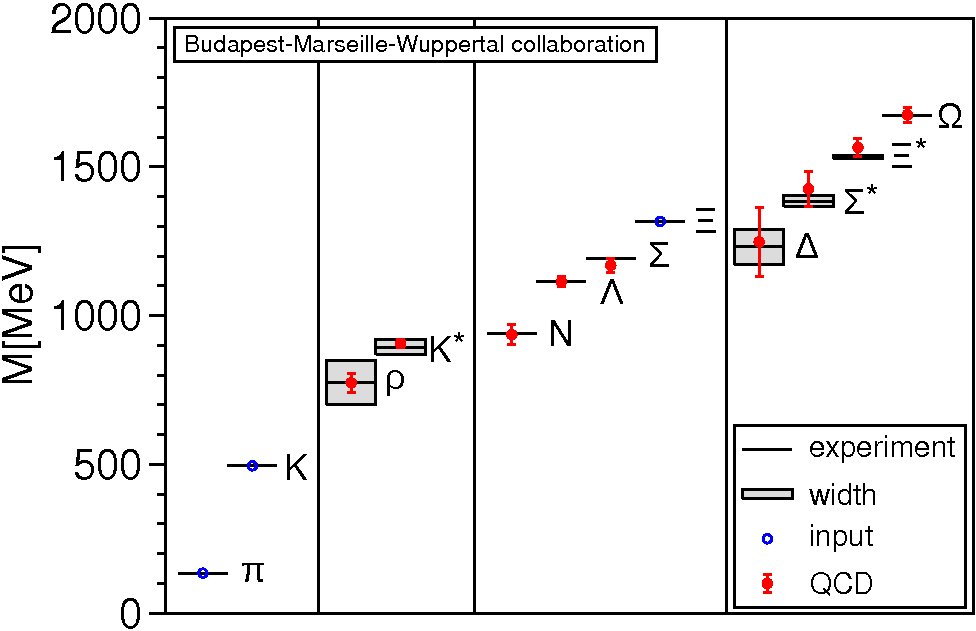} 
   \caption{Spectrum of low-lying mesons and baryons, calculated from first principles in quantum chromodynamics, compared to experiment \cite{qcdSpectrum}.  $\pi$ and $K$ mesons masses are used to fix the light quark and strange quark masses, respectively, and  $\Xi$ to fix the overall scale.   ``N'' denotes nucleon.   This result reveals the origin of the bulk of the mass of standard matter. }
   \label{fig1}
\end{figure}

Note that {\it many\/} distinct localized organizations of energy, or ``particles'', with different masses, arise here as self-consistent solutions of the nonlinear back-reaction equations.

The Lorentzian perspective on the origin of mass in QCD, for all its historical resonance, is perhaps less straightforward than an alternative, simpler yet no less beautiful perspective.   Once we admit that the fields of QCD can conspire to produce quasi-stable, localized concentrations of energy, then special relativity tells us that these entities will behave as particles with mass
\begin{equation}
m ~=~ E/c^2
\end{equation}
where $E$ is the total field energy of the configuration in a frame where its center-of-energy is at rest or, more formally, where its total momentum vanishes.

Figure 1 speaks eloquently for itself.   In view of this triumph, it is appropriate to consider its conceptual roots closely.   A good entry into the discussion is to ask: What is the input, that underpins such impressive output?   

Decades of intense theoretical work have reinforced and indeed strengthened the intuition, originally based on perturbation theory, that 3+1 space-time dimensional quantum field theories, which are the only theories combining special relativity and quantum theory that are known to be consistent, are highly constrained.   The source of this constraint is the possibility of ultraviolet divergences.    In more detail: Special relativity appears to demand that the interactions be local; local interactions bring in field modes with arbitrarily large energy and momenta; coupling to quantum fluctuations in those field modes can integrate up to become so large as to disrupt the expected ({\it e. g}. finite, relativistically invariant) behavior of physical process in the theory, rendering it incalculable and effectively nonexistent \cite{qftReview}.    Nonabelian gauge theories can avoid that fate, because asymptotic freedom weakens the couplings of the dangerous modes.    That happy result only occurs, however, for the minimal gauge invariant couplings.   One cannot tolerate, for example, anomalous color magnetic moments, for either quarks or gluons.    

For this reason QCD, a nonabelian gauge theory based on the gauge group $SU(3)$ and triplet quarks, supports very few parameters \cite{qcdReview}.   One can specify a mass for each quark field, and one overall coupling strength; nothing else is allowed \footnote{Here for once I'll pass over in silence the saga of the $\theta$ term.}.   

The mass parameters associated with quark fields do not correspond directly to the mass of any physical particle, since quarks famously do not occur as individual particles, but only confined within more complicated composites.   We can, however, infer how energetic quarks propagate over very short times, by reconstructing them from the jets they induce in high-energy collisions.    That propagation is of course affected by the quark's mass, and in principle allows its measurement.   This technique is the only method by which the top quark mass $m_t$ can be measured, and it also allows access to the bottom and charm masses $m_b, m_c$.   $m_b$ and $m_c$ can also be inferred from the masses of bottomonium and charmonium resonances, which are fairly well described as non-relativistic $b\bar b$ and $c \bar c$ bound states, respectively.   The light quark masses $m_s, m_d, m_u$ can be inferred from hadron spectroscopy; the masses of the low-lying pseudoscalar mesons $\pi, K, \eta$ in particular depend sensitively upon $m_s, m_d, m_u$.     

From the perspective of QCD dynamics, the quarks divide into two classes, the light quarks $u, d, s$ and the heavy quarks $c, b, t$.   The reason for this division is that QCD dynamics is characterized by another scale with dimensions of mass, to which quark masses can be compared.   This is the scale $\Lambda_{\rm qcd}$ associated with the running, as a function of momentum transfer, of the dimensionless coupling $g(Q)$.   For $Q \lesssim \Lambda$ the coupling is strong, and fluctuations in the gluon field dominate the dynamics.   Hence the effects of {\it quark masses\/} with $m_q \ll \Lambda$ will, ordinarily, be dominated by gluon fluctuations, and will have limited impact on the underlying dynamics; while {\it quarks\/} with $m_q \gg \Lambda$ will ordinarily have limited impact on the gluon dynamics, since the fluctuations are insufficient to produce heavy $q\bar q$ pairs.    Phenomenologically $m_u, m_d \ll \Lambda \ll m_c, m_b, m_t$, while $m_s \lesssim \Lambda$.   

Based on this dichotomy it is amusing and instructive to consider an idealization of QCD which I call QCD Lite, wherein one takes $m_u, m_d, m_s \rightarrow 0$ and $m_c, m_b, m_t \rightarrow \infty$.    In this idealization the heavy quarks effectively decouple, and we arrive at the theory of three flavors of massless quarks.  In view of the preceding paragraph, one might anticipate that the masses of hadrons in QCD Lite closely resemble those of realistic QCD, especially for hadrons free of strange quarks (such as protons, neutrons, $\Delta$ baryons, $\pi$ and $\rho$ mesons, and their rotational excitations).   That expectation can be checked against accurate numerical simulations, and it proves out.  

In this way we are led to a most remarkable conclusion.  QCD Lite is a theory containing {\it no\/} explicit mass parameters; its only dimensional scale is set by the coupling evolution.   Yet it supports a rich spectrum of massive excitations, notably including protons and neutrons, that closely resembles the spectrum of QCD proper.   Thus ``mass without mass'' from QCD dynamics accounts for almost all of the mass of protons and neutrons - and therefore, according to the preceding analysis, almost all of the mass of what we ordinarily consider matter.

\section{Masslessness and Symmetry}

The question of the origin of mass acquires a new level of interest through its deep connection with the symmetry of physical law, for in some important cases {\it massless\/} fields (i. e., fields whose quanta are massless particles) support forms of symmetry that are inconsistent with non-zero mass.     In such cases, the origin of mass gets tied up with the question of dynamical symmetry breaking.

\subsection{Scale Symmetry}

The simplest mass-forbidding symmetry is scale invariance.   In a relativistic field theory it is appropriate to use $c$ as the measure of velocity and $\hbar$ as the measure of action and $c$, so  $\hbar = c =1$.  In such units mass has dimensions of inverse length:
\begin{equation}
M ~=~ \frac{\hbar }{cL} ~\rightarrow~ \frac{1}{L}
\end{equation}
Symmetry under scale transformations $x \rightarrow \lambda x$, which involves $L \rightarrow \lambda L$, is therefore incompatible with the existence of any invariant mass parameter.

\subsection{Chiral Symmetry}

Relativistic spin-$\frac{1}{2}$ fields $\psi$ are said to be left-handed if they satisfy the equation 
\begin{equation}
\Pi_R \psi ~\equiv~ \frac{1 + \gamma_5}{2} \psi ~=~ 0
\end{equation}
and right-handed if they satisfy
\begin{equation}
\Pi_L \psi ~\equiv~ \frac{1 - \gamma_5}{2} \psi ~=~ 0 
\end{equation}
Because the projection operators satisfy $\Pi_L + \Pi_R = 1$, we have $\Pi_L \psi = \psi$ for a left-handed field and $\Pi_R \psi = \psi$ for a right-handed field.   
Fields with a definite handedness, left or right, are said to be chiral.  Minimal chiral fields are spinor fields with two independent components.  

The quanta of a free left-handed field carry negative helicity -- that is, their spin in the direction of motion is $-\frac{1}{2}$.    That condition can only be relativistically invariant for particles that move at the speed of light -- otherwise, one can overtake the particle and reverse the helicity.    So chiral fields describe massless quanta.  Massive spinor fields must contain both left- and right-handed components.   In a Lagrangian formulation, the mass term for free quanta takes the form 
\begin{eqnarray}\label{massTerms}
-{\cal L}_m ~&=&~ =  m \bar \psi \psi \nonumber \\
~&=&~ m (\overline {\Pi_R \psi} \Pi_L \psi + \overline{\Pi_L \psi} \Pi_R \psi ) \nonumber _\\
~&\equiv&~ m (\overline {\psi_R} \psi_L + \overline{\psi_L}{\psi_R} )
\end{eqnarray}
Note that a phase transformation on the $\psi_R$ field, $\psi_R \rightarrow e^{i\theta} \psi_R$, does not leave the right-hand side of Eqn.\,(\ref{massTerms}) invariant.  Looking at the effect of this transformation, we also see that it is natural to contemplate complex values of mass terms, in the generalized form
\begin{equation}\label{complexMass}
-{\cal L}_\mu ~=~ ( \mu^*  \overline {\psi_R} \psi_L + \mu \overline{\psi_L}{\psi_R} ) ~=~ {\rm Re}\,\mu \, \overline {\psi} \psi + i{\rm Im}\, \mu \, \overline {\psi} \gamma_5 \psi
\end{equation}
 
Candidate symmetries can act on left- or right-handed fields separately.  Thus for example if we have two spinor fields with both left- and right-handed pieces, we can contemplate, consistent with relativistic invariance, $U(2)_L \times U(2)_R$ transformations among them; but for massive fields the possible symmetry is reduced to $U(2)_{L+R}$, because the the left- and right- handed components must transform together.   Symmetries that distinguish left- from right-handed quanta, and thus force them to be massless, are called chiral symmetries.  

\subsection{Gauge Symmetry}

Gauge symmetry applies to the quantum fields that describe spin-1 bosons.   It has a beautiful geometrical interpretation, but here I would like to emphasize some more directly physical issues which relate closely to the question of mass and the special status of $m=0$.  

By analogy with chiral symmetry for spin $\frac{1}{2}$, just discussed, one might seek to formulate theories of spin-1 fields whose quanta are only left-handed, or only right-handed; or in other words, only definite circular polarizations.  Those quanta would necessarily be massless, for the same reason we discussed in the spin $\frac{1}{2}$ case.    Consistency requirements related to the CPT theorem require that both helicities be present, but apart from that subtlety gauge symmetry is precisely the implementation of this idea.   A vector field $A_\mu$ contains longitudinal polarizations, in addition to the desired circular polarizations.    Gauge invariance insures that the longitudinal polarizations have no physical effect, since it requires invariance of the theory under addition of a longitudinally polarized field:
\begin{equation}\label{gaugeInvariance}
A_\mu ~\rightarrow~ A_\mu + \partial_\mu \Lambda
\end{equation}
with $\Lambda$ an arbitrary space-time function. 
(Here for simplicity I suppress the intricacies of nonabelian gauge symmetry.)    

Whereas conventional symmetries imply relations among physical processes, gauge symmetry is an essentially mathematical statement, about theory formulation.   Gauge symmetry asserts the absence of degrees of freedom that change under its transformations, rather than, as for non-gauge symmetries, invariant relationships among physical degrees of freedom.  In quantum theory, it is implemented by constructing the physical Hilbert space through projection onto gauge-invariant states.   The formalism of lattice gauge theory is manifestly gauge invariant, implementing projection by direct averaging over the gauge group\footnote{It is not known, however, how to formulate theories with gauged chiral symmetries, including notably the full standard model (as opposed to QCD and QED), in that framework \cite{fermionDoubling}.} \cite{latticeGaugeTheory}.  Conventional perturbation theory works instead by using the gauge freedom to put the external $A_\mu$ field in a canonical form -- ``gauge fixing'' -- while allowing off-shell fluctuations, not subject to projection, at intermediate stages in its calculations.   One must then work to show that the canonical form carries through to the final result, so that the desired projection is consistent.   This can raise challenging technical issues, especially with regard to regularization and renormalization.  It can even go wrong, leading to so-called gauge anomalies.   Failure to implement a consistent projection is highly problematic, specifically because the sign of covariant quantum-mechanical amplitudes involving $A_0$ and the spatial $A_i$ will be opposite, leading to assignment of negative probabilities.   

Gauge symmetry is a central postulate of the standard model.  It is appropriate to ask whether it is a completely independent postulate, or tied up with other fundamental principles.  Special relativity and quantum mechanics inevitably lead to local quantum field theory, but non-trivial quantum field theories of this kind are notoriously difficult to construct, due to ultraviolet divergences.   Perturbation theory with massive vector particles involves the propagator
$$
\frac{g_{\mu \nu} - \frac{k_\mu k_\nu}{M^2}}{k^2 - M^2}
$$
near mass shell (so in the rest frame the time-like polarization is projected out).   This has miserable ultraviolet behavior, being completely undamped as $k \rightarrow \infty$.   For massless vectors with gauge symmetry, on the other hand, we can work in Fermi-type gauges where the propagator assumes the form
$$
\frac{g_{\mu \nu} - \alpha \frac{k_\mu k_\nu}{k^2}}{k^2}
$$
with much better large-$k$ behavior.    The renormalization program delivers on this promise, by giving a procedure for calculating physical amplitudes as a power series in the renormalized coupling, with finite coefficients.   That is not good enough, however, since the series can fail to converge; we must investigate the non-perturbative behavior.   In principle the simplest, and in practice the most powerful, approach to constructing interacting quantum field theories is to introduce some sort of cutoff to regulate the ultraviolet behavior, calculate within the regulated theory, and remove the regulator by a limiting procedure, while holding some physical observable fixed.   In general this approach will fail, because as one raises the cutoff, one (re)introduces significant effects from ever higher-frequency fluctuations, and no definite limit can emerge.   In asymptotically free theories, however, the high-frequency fluctuations decouple, and this constructive strategy works.   Since asymptotic freedom is an issue concerning the ultimate weakness of couplings, it can be investigated perturbatively.  Famously, one finds that asymptotic freedom is a feature unique to (nonabelian) gauge theories.  Thus gauge invariance -- indeed, nonabelian gauge invariance -- is at least helpful, and probably necessary, for the harmonious marriage of special relativity with quantum mechanics.    Thereby gauge invariance, and the consequent existence of fields with $m=0$ vector quanta, appears deeply inherent in the nature of things.

\section{Mass and Symmetry Breaking in QCD}

QCD features all three of the mass-forbidding symmetries just discussed, yet manages to produce the mass of matter.  This is especially clear and dramatic in realistic QCD's quasi-realistic cousin QCD Lite, whose formulation contains no quantity with units of mass or length.   In this subsection I'll elaborate a bit on the symmetry-breaking aspects of mass generation in QCD.   Aside from their intrinsic interest, these considerations shed interesting light on the Higgs mechanism and the Higgs particle, as will appear below.

\begin{itemize}
\item {\it Scale Symmetry}: The classical scale invariance of QCD Lite, which characterizes its Lagrangian formulation, is violated by quantum fluctuations, leading to scale dependence of its dimensionless coupling.  Because the theory is nonlinear, larger-scale fluctuations are influenced by fluctuations on smaller scales, and that influence alters their properties.   That is the simple physical reason why scale symmetry, while valid in the classical formulation of the theory, is necessarily broken upon quantization.  Thus this symmetry barrier to mass generation falls.   

Though it is literally false, scale symmetry involves important truths.   First of all, it becomes true asymptotically, and thereby governs the high-energy behavior of many physical processes.  Two other profound consequences relate directly to our overarching theme, the origin of mass.   Since the scale invariance of classical QCD Lite would make it impossible for nucleons to have mass, we see that {\it the bulk of the mass of matter is essentially quantum-mechanical in origin}.    Scale invariance is replaced by an emergent relationship between coupling strength and scale, whereby neither coupling strength nor scale can be freely varied, but variation in one can be compensated by variation in the other.   In place of scale invariance of the theory, we have what Sidney Coleman called ``dimensional transmutation'': Different values of the coupling strength do not yield a family of essentially different theories, but rather correspond to a unique theory equipped with different units of length (or mass).   Thus {\it the bulk of the mass of matter arises from a unique, parameter-free theory}.   

\item {\it Gauge Symmetry}: Gauge invariance states that the physical Hilbert space contains only singlet states.   Since all physical states are gauge singlets, gauge invariance has no direct implication for the physical spectrum, even though it constrains the mass parameter of gauge {\it fields}, which are used in constructing the theory, to vanish.  Putting it more simply and physically, though more loosely: The nominally massless plane-wave gluons are confined, while the ``cavity modes'' associated with allowed, finite-extent configurations have a non-zero minimum frequency.    

It seems appropriate to remark here that confinement of color gluons, and quarks, is a very natural consequence of gauge symmetry, since those quanta are {\it not\/} gauge singlets.   From this perspective it is the deconfined phases of QED and electroweak theory, which appear so natural in perturbation theory, that appear paradoxical.    

\item {\it Chiral Symmetry}:  We might also invoke confinement to explain away massless quarks.   Chiral symmetry has profound implications for the spectrum, nevertheless.   

To simplify the discussion, I will ignore the strange quark, and consider a 2-flavor version of QCD Lite.   Representations of chiral $SU(2)_L \times SU(2)_R$ are of two kinds.  We can have massless fermion multiplets, as for free fermions, or massive multiplets with degenerate states of opposite parity.   Neither alternative is a feature of the observed hadron spectrum.  

Fortunately, in QCD chiral symmetry is broken, dynamically (or as we say ``spontaneously''),  through formation of a quark-antiquark condensate in the ground state.   We can write 
\begin{equation}\label{condensate}
\langle G(U) | \overline {q_{j R}} q^k_L | G(U) \rangle ~=~ v U^k_j
\end{equation}
where the subscripts $L, R$ denote left- and right-handed chirality, $j, k$ are flavor indices, $v \neq 0$ is the magnitude of the condensate, and $U$ is a unitary matrix with determinant unity.   Any fixed choice of $U$, independent of space-time coordinates, gives a possible ground state $| G(U) \rangle$, and these states are all orthogonal, but energetically degenerate.  Slow space-time variation of $U$ creates low-energy modes that, following Nambu and Goldstone, we identify as representing pseudoscalar mesons. 

In addition we might consider space-time variation in the {\it magnitude\/} of $v$.   Thus we consider imposing (with $U \rightarrow \delta$)
\begin{equation}\label{condensate}
\langle | {\bar q}_{j L} q^k_R | \rangle ~=~ \bigl( v + \sigma (x, t) \bigr) \delta^k_j
\end{equation}
on field histories, and forming an effective action.  In the language of the old $\sigma$ model \cite{nambu}, \cite{sigmaModel} we represent these excursions in and around the vacuum manifold as
\begin{equation}
v(x, t) U(x, t) ~=~ (v + \sigma) \exp \, (i \vec \tau \cdot \vec \pi)
\end{equation}
with Pauli matrices $\vec \tau$ and a triplet of pion fields $\vec \pi$.

Since variations in $\sigma(x, t)$ take us outside the vacuum manifold, we should not expect them to have vanishing energy, even in the limit of long wavelength.    It may be possible to associate an observed excitation in QCD, the broad scalar meson resonance $\sigma(500)$ around 500 MeV, with variations of this kind.   

For later comparison with the minimal Higgs sector, an alternative notation for $v(x,t)  U (x, t) $ is significant.   We can write
\begin{equation}
vU ~=~ \left(\begin{array}{cc}\phi_0 & \phi_1 \\ -\phi_1^* & \phi_0^* \end{array}\right)
\end{equation}
with complex scalar fields $\phi_0, \phi_1$.   These form a doublet under multiplication by $SU(2)$ matrices $V$ acting as $vU \rightarrow vUV$, or in other words under $SU(2)_L$.  In this notation the vacuum expectation value is $v = \langle {\rm Re} \, \phi_0 \rangle$ and the $\sigma$ meson field, for small fluctuations, is $\sigma \approx  {\rm Re} \, \phi_0 - v$.   This corresponds precisely, as we'll see, to the minimal Higgs particle.

\end{itemize}

\bigskip

\section{Summary: The Origin of Mass for Standard Matter}

\bigskip

By way of summary, some major statements concerning the mass of standard matter:

\begin{enumerate}
\item In classical mechanics Mass is an irreducible, defining property of matter.   In modern physics there is no equivalent fundamental concept.  Mass in the original, Newtonian sense appears as an approximate, emergent property of matter.  
\item There is a contingent but clear conceptual path leading from qualitative aspects of the quantum dynamics of elementary quanta to Newtonian mass as an approximate property of bulk matter.   Thus the issue of the origin of mass, for standard matter, reduces to understanding the properties of the relevant elementary quanta.  
\item The elementary quanta that give quantitatively important contributions to the mass of standard matter are color gluons, $u$ and $d$ quarks, and to a much lesser extent electrons and photons.  Everything else is basically negligible.   (The contribution of the $s$ quark is tricky to separate from the gluons, since most of its effect can be absorbed into a renormalization of the coupling.)   
\item The equations for massless fields exhibit several kinds of enhanced symmetry: scale invariance, chiral symmetry, gauge symmetry, that are spoiled by non-zero mass.   By first assuming those symmetries, and then explaining how they are transcended, we can claim sharp insight into the {\it origin\/} of mass.   
\item We have superb calculations accounting for the origin of nucleon masses, and the masses of hadrons generally, based on QCD.  A highly symmetric, parameter-free version of QCD that incorporates scale, chiral, and gauge symmetry already provides a good approximation to the relevant hadronic spectrum, and accounts for most of the mass of ordinary matter.   Within this framework, we find very concrete understanding of how each mass-forbidding symmetry is transcended.  
\end{enumerate}

I would like to emphasize that this theory of the origin of the mass of standard matter makes no reference to symmetry breaking in the electroweak sector, nor to the Higgs condensate, nor of course to the Higgs particle.    Those are important subjects, but they are different subjects, to which we now turn.


\section{Mass and Symmetry Breaking in Superconductivity}

Photons in empty space have zero mass, for a profound reason, connected to (unconfined) gauge symmetry.   Yet the equations for photons inside superconductors describe a massive particle.  That is a deep interpretation of the Meissner effect, implicit in the work of London and of Landau and Ginzburg.   In the context of superconductivity, those authors were primarily focused on the classical behavior of the electromagnetic fields and their effect on external probes, not the appearance of field quanta to imaginary observers immersed in the superconducting material.  Nevertheless their equations, analyzed from that perspective, unambiguously correspond to a massive photon.  

The Landau-Ginzburg framework is perfectly adapted to re-interpretation and generalization.  In particular, the fields that appear in their construction can be promoted to relativistic quantum fields, and the gauge symmetry can be taken to be nonabelian.   Those generalizations, and the interpretation of the field quanta as massive vector particles, were developed in the work of several authors (including Higgs \cite{higgs}; Brout and Englert \cite{broutEnglert}; Guralnick, Hagen, and Kibble \cite{GHK}; 'tHooft \cite{'tHooft}) whose separate contributions I will not attempt to disentangle here.   The essence of the matter can be captured in a few strokes.  We consider the interaction of the electromagnetic (or more general gauge) field with a complex (or non-singlet) scalar field described by the relativistic Lagrangian
\begin{equation}
{\cal L} ~=~ - \frac{1}{4} F_{\mu \nu}F^{\mu \nu} + (\nabla_\mu \phi)^*(\nabla^\mu \phi) - V(\phi^* \phi)
\end{equation}
where $V$ is an invariant potential and 
\begin{equation}
\nabla^\mu \phi ~=~ \partial^\mu \phi -i q A^\mu \phi
\end{equation}
is the covariant derivative.   If we suppose that $V$ is minimized away from the origin
\begin{eqnarray}
V(v^2) ~&=&~ V_{\rm min.} \nonumber \\
v ~&\neq&~ 0 
\end{eqnarray}
and substitute this value into the covariant derivative terms, we find a term
\begin{eqnarray}
{\cal L } ~&\supset&~ m^2 A_\mu A^\mu \nonumber \\
m^2 ~&=&~ q^2 v^2 
\end{eqnarray} 
that corresponds to the indicated (mass)$^2$ for the vector field.   

After Bardeen, Cooper, and Schrieffer, we understand that in superconductors the $\phi$-field represents the density of Cooper pairs, with $q = 2e$. 

\subsection{Gauged Broken Symmetry, NOT Broken Gauge Symmetry}

The common usage ``broken gauge symmetry'' to describe superconductivity and its electroweak analogue is unfortunate.    As I've already had occasion to recall, local gauge invariance cannot be broken, since it is a statement about theory construction.    Yet the common usage was not chosen whimsically, or perversely; clearly there is something in it.  What?  

In a weakly coupled gauge theory, we are invited to consider the theory we would get without the gauge fields, and then add in the gauge interactions as a perturbation.   This is what we actually do, in both superconductivity and in the minimal electroweak theory.   In the absence of gauge interactions, we have a model of broken global symmetry; then we couple in the gauge fields, and compute the spectrum perturbatively, finding massive gauge bosons.    

(Optional digression: An amusing point is that since ordinary superconductivity arises, ultimately, primarily from electromagnetic forces, the concept of ``turning off'' the gauge field in that context appears paradoxical.   Within the Landau-Ginzburg framework, of course, no paradox arises.  The potential has a life of its own, with its dependence on the underlying microscopic forces suppressed.    Presumably some separation of the instantaneous Coulomb interaction, which should be treated nonperturbatively, from the perturbative propagating transverse field could help here, or a renormalization group derivation of effective interactions whose strengths are regarded, for conceptual purposes, as independently adjustable parameters.)

Thus we should properly speak of gauged broken symmetry, as opposed to broken gauge symmetry.   That distinction might appear pedantic, but it clarifies a profound conceptual unification, as will appear momentarily.

\section{Mass for W and Z Bosons}

In the context of SU(2)xU(1) electroweak gauge theory we get a good account of the origin of $W$ and $Z$ boson masses and mixings along these lines, using a condensate having the quantum numbers of a component an $SU(2)$ doublet $\phi$ with hypercharge $Y/2 = 1/2$.   The details are textbook material \cite{chengLi}, which I'll forego here.   It is appropriate to mention a conceptually important feature of the derivation, however: The new structure it introduces into the theory is quite minimal.   One must specify the gauge quantum numbers of the condensate, and its magnitude; that is all.   Since the ratio of output to input is large and impressive, I think it is fair to claim that this mechanism greatly illuminates the origin of mass, {\it for the W and Z bosons}.    The historical sequence of events was likewise impressive: Since both the strength of the $SU(2)\times U(1)$ gauge couplings and the Fermi constant $G_F$, which sets the mass scale, had been measured, it was possible to form precise expectations for the masses and properties of the $W$ and $Z$ bosons prior to their experimental discovery.   

The known particles of the standard model, that is the quarks, leptons, and gauge bosons, do not by themselves produce the required condensate.  The quark-antiquark condensate of QCD actually has the right quantum numbers for the job, but it is far too small quantitatively.     To produce the required dynamics, the simplest procedure is to introduce an elementary scalar doublet with the required quantum numbers.   This can be done using a direct extrapolation of the Landau-Ginzburg Lagrangian.   An important practical virtue of this construction is that it can be carried out in the framework of weak coupling, renormalizable quantum field theory, provided we restrict $V$ to the form 
\begin{equation}\label{higgsPotential}
V(\phi^* \phi) ~=~ - \mu^2 \, \phi^* \phi + \lambda \, (\phi^* \phi )^2
\end{equation}
In this was we get quite a definite theory, whose consequences can be calculated accurately.   I will discuss it under the heading ``The Minimal Model'' in the following Section.  

Of course, as a logical matter reality need not be so simple.   The condensate might arise from several distinguishable contributions, {\it e. g}. from several distinct doublets, or from other representations, or from composites beyond the known, small quark-antiquark portion.   

\subsection{Continuity of Confinement and the Higgs Mechanism (and Superconductivity)}

I've discussed two mechanisms for explaining the origin of mass, that operate in different contexts -- QCD and confinement for hadrons, and the Higgs mechanism for $W$ and $Z$ (which is furthermore closely related to superconductivity).  Superficially those mechanisms appear quite different, but at a fundamental level they are essentially the same.   Indeed, as we imaginatively turn up the $SU(2)$ gauge coupling, keeping the potential $V$ fixed, we will go from dynamics most conveniently described as ``gauged broken symmetry'' to dynamics most conveniently described as ``confinement''.  But as long as the non-photonic spectrum remains massive throughout this thought-experiment there can be no sharp transition.     

If we ignore the gauge couplings of the standard model altogether, we find that the minimal Higgs sector has precisely the same structure as the old $\sigma$ model (in its linear incarnation), an effective theory for low-energy pion physics, in the idealization of exact chiral $SU(2)_L\times SU(2)_R$.  The ungauged Higgs sector likewise features an $SO(4) \sim SU(2)\times SU(2)$ symmetry, that is spontaneously broken to the diagonal subgroup.   In $\sigma$ model language, this spontaneous breaking produces three massless Nambu-Goldstone bosons, the pions, plus the $\sigma$ meson.   When we turn on the gauge couplings the pions combine with the gauge fields, providing the longitudinal modes for now massive gauge bosons.   We see, from this correspondence, that the Higgs particle $H$ is quite closely related, conceptually, to the $\sigma$ meson of QCD.   Indeed, if the Higgs condensate were dominated by its quark-antiquark contribution, the $\sigma$ meson would {\it be\/} $H$.    

The concepts here briefly reviewed play a prominent role in technicolor model building, where hypothetical new higher-scale analogues of QCD are introduced to produce a dominant contribution to the electroweak condensate.   Whatever the fate of such models, the deep connection between our two best ideas for understanding the origins of mass in different contexts remains a satisfying insight.

\section{Mass for Quarks and Leptons}

In the $SU(2)\times U(1)$ electroweak theory left- and right- handed fermions couple differently.  Therefore mass terms, which connect left and right, are not gauge invariant.    
Fortunately, the same doublet condensate that works for $W$ and $Z$ can also serve here, to connect the differently handed chiral fermions. 
In detail, for quarks:   

\begin{equation}\label{fermionYukawa}
{\cal L } ~\supset~ -y^{jk}\, {\bar U}_{j} Q_{\alpha k}\, \phi^\alpha - z^{jk} {\bar D}_{j} Q_{\alpha k} \, \epsilon^{\alpha \beta} \, \phi^*_\beta + {\rm h.c}.
\end{equation} 

When the appropriate component of $\phi$ acquires a vacuum expectation value, and we insert that value into Eqn.\,(\ref{fermionYukawa}), then we obtain a quadratic expression in the quark fields.   By some complicated field redefinitions (even more complicated, if we take into account possible family structure in the kinetic Dirac Lagrangian) we can diagonalize that expression, to obtain physical masses of the independently propagating quanta.    The relative rotations necessary for diagonalization encode the weak mixing angles, or Cabibbo-Kobayashi-Maskawa (CKM) matrix.   In this construction parameters proliferate, unconstrained by theoretical understanding.  

There is a similar construction for leptons, with the additional complication that (in the minimal model) neutrino masses arise from non-renormalizable dimension five interactions, that are quadratic in $\phi$.  

To call this mess ``the origin of mass'', even for the elementary quanta, is to aim too low.  The measured masses and mixings of quarks and leptons are in need of elucidation, but this construction manifestly shirks the burden of explaining them.  It is, at best, an accommodation.    Similarly, the quantity $\mu$ that appears in Equation\,(\ref{higgsPotential}), which has dimensions of mass, appears as a free parameter, and is no way elucidated or explained.  Yet this $\mu$ is intimately related to the mass-scale of the condensate and it governs the physical mass of the classic Higgs particle itself quite directly.

\section{The Message of Higgs Particle Phenomenology}

\subsection{The Minimal Model}

Now let me turn, at last, to the recent excitement in Higgs particle phenomenology.  

For the sake of organizing the discussion, I will first consider as default hypotheses:
\begin{itemize}
\item First hypothesis: The minimal version of the standard model, with just one doublet, is a good approximation to reality (in the relevant energy range).
\item Second hypothesis: All indications from the relevant LHC (and LEP,  Tevatron, ...) experiments are correct.
\end{itemize}
The immediate question then becomes: Do these hypotheses hang together?

As mentioned previously, a virtue of the first hypothesis is that it gives us a definite, quantitative, one-parameter theory, that covers an enormous range of phenomena.  Despite years of hard effort to poke holes in that theory, so far it has withstood every test.   Most of its predictions, of course, are not very sensitive to the value of the parameter $m_H$, which after all affects only virtual $H$ particles, since the $H$ mostly interacts feebly, especially with the light quarks, photons, electrons, and gluons that we have easiest access to.   $H$ exchange has small but measurable effects on the $W$ and  $Z$ masses, and on some other very accurately measured electroweak parameters.    The constraint from these estimates, according to the LEP Electroweak Working Group, is shown in the following Figure: 
\begin{figure}[h!] 
   \centering
   \includegraphics[scale=.3]{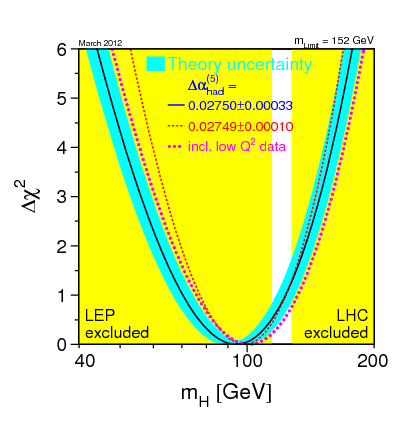} 
   \caption{Constraint on Higgs particle mass, based on analysis of radiative corrections, as analyzed by the LEP Electroweak Working Group \cite{yellowBand}.}
   \label{fig2}
\end{figure}


The message that emerges from these exquisite calculations and experiments, that involve every part of the standard model and deep, intensive use of quantum field theory, is that 
agreement with the minimal model is possible, but only if $m_H$ is below 140 GeV or so.   Since the LEP direct search excluded $m_H \lesssim 115$ GeV, one had only a narrow window for consistency, even prior to the LHC work.

The situation at LHC is developing rapidly, so this is a dangerous time to interpret the situation.   Nevertheless ... !

As we see in the next Figure, there are significant hints of an $H$ signal at $m_H \approx 125$ GeV.   
 \begin{figure}[h!] 
   \centering
   \includegraphics[scale=.45]{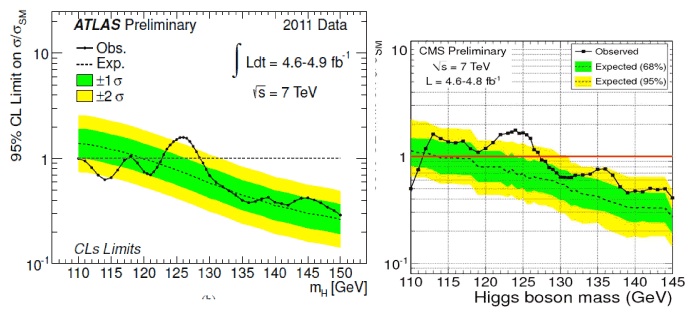} 
   \caption{Preliminary hints of signal consistent with minimal standard model Higgs particle, $m_H \approx 125$ GeV \cite{higgsHint}.}
   \label{fig3}
\end{figure}

The signal recorded there has been distilled from a combination of several channels, and the details are rather complicated.   I'd like briefly to discuss one (dominant) contribution to the signal, which is particularly clean and has some special theoretical interest.   It is the process sketched here: 
 \begin{figure}[H] 
   \centering
   \includegraphics[scale=.2]{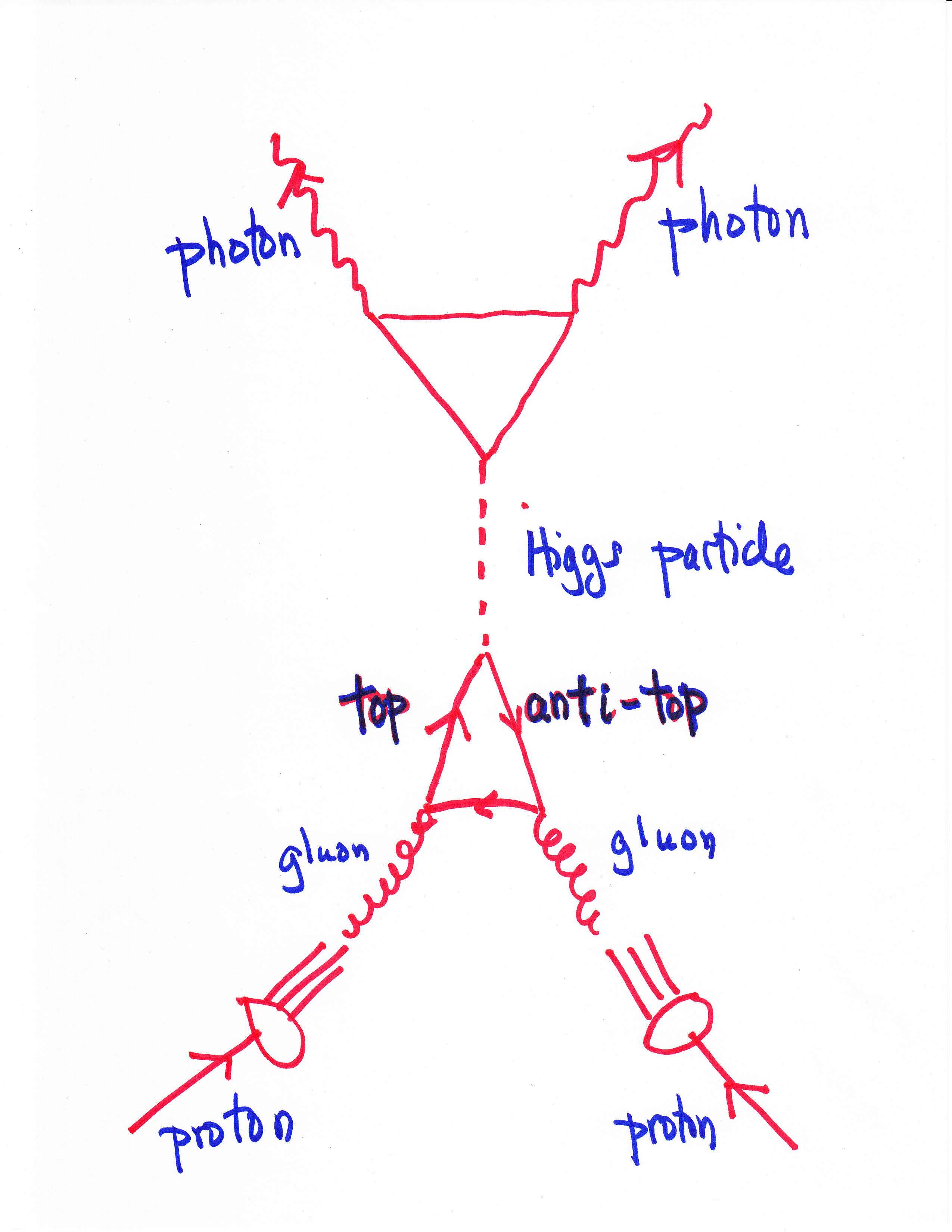} 
   \caption{Stylized Feynman graph illustrating the production of a Higgs particle through gluon fusion and its decay through photon fission.  Time advances upwards.}
   \label{fig4}
\end{figure}
%

The experimental signature is a resonant enhancement in $\gamma \gamma$ production, when the invariant mass of the photons matches $m_H$.   One sees that this process is basically a doubled version of the coupling of $H$ to gauge bosons, displayed here:
 \begin{figure}[H] 
   \centering
   \includegraphics[scale=.25]{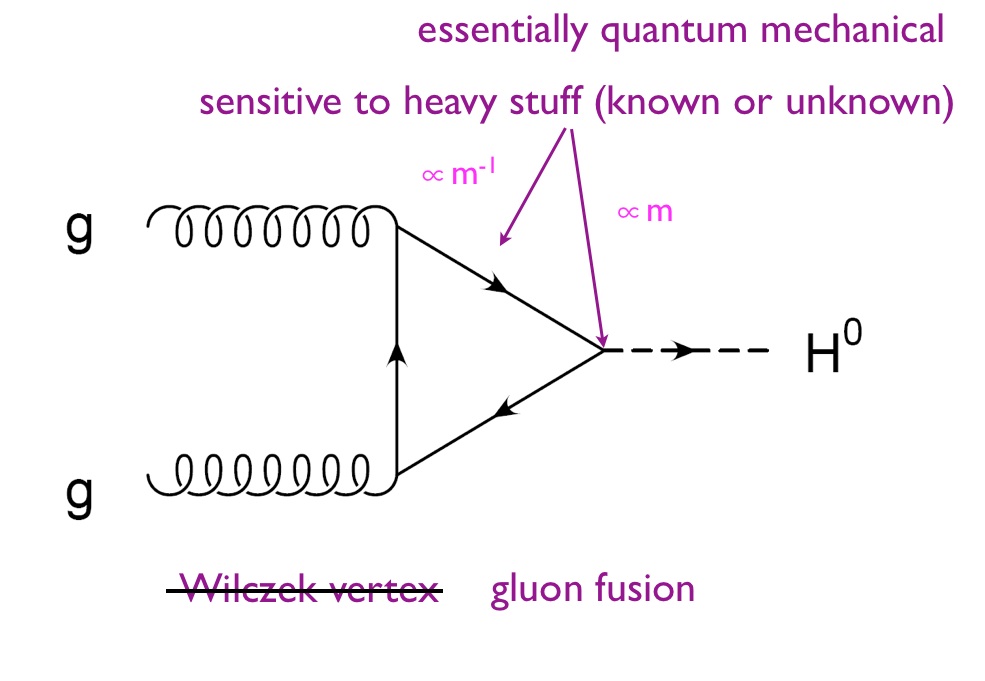} 
   \caption{Origin of the effective coupling between Higgs particle and gluons.   The annotations are explained in the text.}
   \label{fig5}
\end{figure}

I'm particularly fond of this coupling, which for a brief shining moment following its discovery \cite{wilczekVertex} was called the ``the Wilczek vertex'', but somehow devolved to ``gluon fusion''.    $H$ has a difficult time coupling to ordinary matter.  Indeed, it prefers to couple to the quanta of heavy fields (or rather, the quanta of fields it couples to thereby become heavy!), while the $u,d,e, \gamma$ and color gluon $g$ fields have tiny or zero mass.    Thus the direct, classical couplings of $H$ to ordinary matter are highly suppressed.   $H$ can, however, communicate with $g$ pairs through a quantum effect, as indicated by a loop graph.   The gluons communicate with virtual top quarks, which in turn communicate with $H$.  

Ordinarily one might expect that processes involving virtual heavy particles are suppressed by inverse powers of the mass, but this process is an exception.  Indeed, by power counting (taking into account gauge invariance, which brings out two external momenta for the gluons) the loop integral converges linearly, and so the large momenta contribute $\propto 1/M$, where $M$ is the mass of the virtual particles.   However the basic, Yukawa coupling at the $H$ vertex is $\propto M$, so the $M$ dependence cancels.   Thus the $Hgg$ coupling gets essentially equal contributions from all sufficiently heavy quarks.  (Quarks with $m_Q \ll m_H$ will not contribute, however, since $m_H$ provides an infrared cutoff.)   

In any case, the main conclusion, as tentative as it is welcome, has to be that our two default hypotheses do indeed hang together.   The simplest, minimal implementation of the standard model, with $m_H \approx 125$ GeV, fits all the facts brilliantly.   

It is appropriate to remark that both the successful theoretical calculation of rates and backgrounds for observable consequences of $H$, and the experimental detection of its rare subtle signals in an extreme, complex environment will be, if confirmed, scientific achievements of the highest order.

\subsection{Beyond the Minimal Model}

The tight fit between experiment and a minimal, weakly-coupled theory is clearly bad news for speculations about new strongly interacting sectors at the electroweak scale, including technicolor or large extra dimensions in most or all of their many variants.   

More interesting, I think, are the implications for low-energy supersymmetry.   Low-energy supersymmetry can be implemented in the framework of weak coupling.  It also features many cancellations among virtual boson and fermion loops.  Thus, despite the large number of superpartners it requires, supersymmetry can hide itself pretty efficiently.   

Lest we forget, let me remind you, with one figure, that there is an excellent, quantitative reason \cite{couplingUnification} to suspect that low-energy supersymmetry is relevant to the description of Nature:
 \begin{figure}[H] 
   \centering
   \includegraphics[scale=.45]{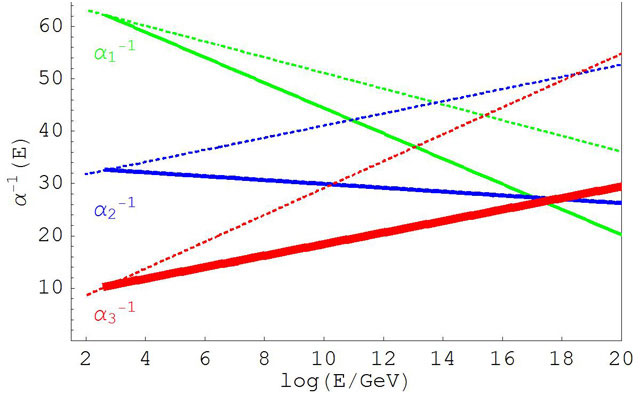} 
   \caption{Quantitative comparison of unification of couplings without (dotted lines) and with (solid lines) contributions from low-energy supersymmetry \cite{unification}.}
   \label{fig6}
\end{figure}
%

If $m_H \approx 125$ GeV, we will have another.  For while the standard model itself has nothing much to say, theoretically, about the value of $m_H$, tasteful implementations of low-energy supersymmetry produce an $H$-like particle whose mass is closely tied to $m_W$ and $m_Z$.   Indeed the value of $m_H$ inferred classically -- i.e. ignoring virtual loops -- satisfies $m_{H, {\rm cl}.} \leq m_Z$, and has long been excluded by experiment.  

Fortunately loop corrections, especially from stop loops, can raise $m_H$.  $m_H = 125$ GeV fits most comfortably with quite heavy ($\sim$10 TeV?) stop masses. 

The idea that the superpartners of quarks and leptons might have such heavy masses had been anticipated on the basis of flavor violation and $T$ (non-)phenomenology.  All current observations are consistent with no additional contributions beyond what arises in the minimal standard model.   Exchange of superpartners, on the other hand, can induce many effects of that kind.  The simplest way to dispose of these dangerous possibilities is simply to suppress the exchanges, by making the superpartners heavy.      

Several theorists have considered this scenario to be natural, and studied it in detail, under the heading {\it focus point supersymmetry} \cite{focusPoint}.   

In detail, supersymmetry requires at least two doublet $\phi$ fields.   The physical spectrum of spin-0 ``Higgs'' particles includes one that mimics the minimal H, plus 2 other neutrals, and a charged $H^{\pm}$.  Of course these particles, like the particles of the minimal standard model, all come with superpartners!

We shall see ... 

\section{Conclusion: Mass in the Universe}

As reviewed here, we have achieved profound insight into the origin of mass for standard matter, and we may be set to crown, with the discovery of the Higgs particle, a compelling account of the origin of mass for $W$ and $Z$ bosons.   Those origins are distinct, though there is an attractive conceptual connection between their mechanisms, and between both mechanisms and superconductivity.   That's the good news.  

The bad news is that nothing in these ideas explains the origin of the mass of the Higgs particle itself, nor do they greatly elucidate the observed complicated structure of quark and lepton masses and mixings, nor the associated physical phenomena of CP violation, neutrino oscillations, ... .  

The ugly news is that the origin of most of the mass in the universe, that in dark matter and dark energy, remains deeply mysterious.    If the dark matter is in axions, the dynamical origin of its mass will be clear, as it arises from known effects in QCD ($\theta$ term), although the magnitude of that mass is tied up with yet another mass scale, the scale $F$ of Peccei-Quinn symmetry breaking.   If the dark matter is some kind of supersymmetric particle, the question of the origin of its mass will be tied up with the larger question of supersymmetry breaking, concerning which there are many speculations, none compelling.  

We've passed some milestones, but the end of the road is not in sight.  

\section{Update: Higgs Particle Discovery}

(Added 18 August 2012.)

On July 4 2012 the ATLAS and CMS experimental groups, reporting work at the Large Hadron Collider (LHC) at CERN, announced the discovery of a remarkable new particle.    Their results have now been codified in scientific papers \cite{atlas} \cite{CMS}.    Perhaps the simplest and most easily interpreted reported signature is an enhancement in $\gamma \gamma$ pair production, over standard model backgrounds, in the invariant mass region $M \approx 125$ GeV.    

Although significant deviations are not yet excluded, the observations so far are close enough to calculated expectations for a minimal standard model Higgs particle, with mass  $M \approx 125$ GeV, that that hypothesis must be considered the default interpretation.   A pre-existing cluster of theoretical and phenomenological arguments, reviewed in earlier Sections, all pointing in the same direction, powerfully reinforce that interpretation.   

Nothing from my earlier discussion needs revision in light of the experimental discoveries (on the contrary, they confirm its logic), so I've left it unchanged.   

Here I'll add some brief but pointed comments about the significance of the discovery in a broader context, especially in its implications for future work.   To avoid possible confusion, in this addendum I will use lower-case $h$ to denote the observed Higgs particle.  (In the literature of supersymmetric phenomenology, it is conventional to use $h$ for the light, standard-model like scalar; one also has a second doublet $H^0, A, H^{\pm}$ of CP even $H^0$ and CP odd $A$ neutral, and charged $H^{\pm}$ spin 0 particles; furthermore $h$ and $H^0$ can mix.) 

Since the rates of many processes involving the Higgs particle $h$ can be measured \cite{peskin}, we can look forward to detailed quantitative comparisons between the observed values of fundamental $h$ couplings and standard model expectations, for several different channels: $WW$, $ZZ$, $\bar t t$, $\bar b b$, $\bar \tau \tau$, $g g$, $\gamma \gamma$, and ``invisible'' (missing energy).    The results will be revealing, since there are serious reasons to anticipate deviations:

\begin{enumerate}
\item {\it mass to mess}: 

As reviewed above (and emphasized already in [22]) the coupling of $h$ to fermions is not deeply rooted in profound principles, and theoretically it is open to easy modification.  Specifically: If the $W$ and $Z$ get part of their mass from a doublet that does not couple to all fermions, then the vacuum expectation value $v^\prime$ associated with $h$ could be smaller than the canonical value, and the corresponding Yukawa couplings $y_{\bar f f} \propto m_f/v^\prime$ larger.   More generally, one can preserve the successful account of $W$ and $Z$ masses and neutral currents in the Standard Model while allowing $h$ to be a mixture of several singlets and doublets, with perturbed couplings. 
 
\item {\it virtual heavies}: 

Also as emphasized already in [22], and reviewed above, the virtual loops that underlie the $gg$ and $\gamma \gamma$ channel can be sensitive to otherwise unknown heavy particles.   (But note that complete cancellation of mass factors between vertex and loop integral, indicated in Figure \ref{fig5}, only occurs for heavy particles whose sole source of mass is coupling to $h$.)  

\item {\it portal; hidden sectors}: 

The Higgs field, because it features the only super-renormalizable couplings in the standard model, is uniquely open to contamination from $SU(3)\times SU(2) \times U(1)$ singlet ``hidden sectors'' \cite{wPrague}.   Such mixings could lead to overall diminution of the production rate, and/or invisible decay channels.   For this reason, the Higgs particle is said to open a portal into hidden sectors.  

\item {\it supersymmetry}: 

Supersymmetry supplies both fodder for $hgg$ and $h\gamma \gamma$ loops, and possibilities for $h-H^0$ mixing.
\end{enumerate}

\section*{Acknowledgements} I'd like to thank Jonathan Feng for helpful comments on the manuscript, and Jesse Thaler for post-discovery orientaton.   This work is supported by the U.S. Department of Energy under cooperative research agreement Contract Number DE-FG02-05ER41360.

\end{document}